\documentclass[prl,aps,twocolumn,preprintnumbers]{revtex4}

\usepackage{graphicx}
\usepackage{dcolumn}
\usepackage{bm}

\begin{document}

\preprint{Optics Express {\bf 13}, 660-665(2005)}

\title{Continuous operation of a one-way quantum key distribution system over installed telecom fibre}

\author{Z. L. Yuan}
\email{zhiliang.yuan@crl.toshiba.co.uk}

\author{A. J. Shields}%
\email{andrew.shields@crl.toshiba.co.uk}

\affiliation{Toshiba Research Europe Ltd, Cambridge Research
Laboratory, 260 Cambridge Science Park, Milton Road, Cambridge, CB4
0WE, UK }

\date{\today}

\begin{abstract}
We demonstrate a robust, compact and automated quantum key
distribution system, based upon a one-way Mach-Zender
interferometer, which is actively compensated for temporal drifts in
the photon phase and polarization. The system gives a superior
performance to passive compensation schemes with an average quantum
bit error rate of 0.87\% and a duty cycle of 99.6\% for a continuous
quantum key distribution session of 19 hours over a 20.3~km
installed telecom fibre. The results suggest that actively
compensated quantum key distribution systems are suitable for
practical applications.
\end{abstract}

\maketitle


Quantum cryptography, or more precisely quantum key distribution
(QKD) \cite{bennett84}, provides a secure way to distribute
cryptographic keys on fibre optic networks, the secrecy of which can
be guaranteed by the laws of quantum mechanics. Most QKD systems
that have been realized so far have encoded the bit information as a
phase delay in an interferometer
\cite{townsend93,townsend94,hughes00,gobby04a,gobby04b,muller97,stucki02,bourennane99,bethune02}.
This is largely due to the availability of high-quality telecom
phase modulators, as well as the fact that phase-encoded qubits are
relatively resilient to decoherence in optical fibres
\cite{bennett92}. Despite this, stabilising the interferometer path
lengths to within the wavelength of the photons is a major technical
problem, currently limiting the usefulness of many applications.

An early long distance demonstration of QKD was based upon two
asymmetric Mach-Zender interferometers (AMZI)
\cite{townsend93,townsend94}. Photons generated by the sender
(Alice) travel through the first AMZI, followed by the optical
fibre, before passing through a second AMZI at the receiver's (Bob)
site.  Optical interference will take place if the phase delay of
Bob's AMZI cancels that of Alice's. Alice and Bob can then perform
QKD using phase modulators in each of the two interfering paths,
provided there is a fixed phase relationship between the two paths
without any modulation.  In practice, maintaining a fixed path
length difference (to within several tens of a nanometer) between
the two AMZIs is very difficult.  In particular, changes in the
ambient conditions, such as the apparatus temperature, cause a slow
drift of the phase difference. Consequently, one-way systems
typically operate for only a few minutes at a time and require
constant realignment \cite{gobby04a, gobby04b}.

The phase drift problem was resolved using an ingenious passive
compensation scheme \cite{muller97,stucki02,bourennane99,bethune02}.
Here the photons make a double pass through a single AMZI, ensuring
passive compensation of any phase drifts occurring over time scales
greater than the propagation time of the photons through the system.
In the 'plug-n-play' scheme \cite{muller97,stucki02}, a bright laser
pulse is divid into two by propagation through the AMZI at Bob and
then transmitted to Alice.  Alice modulates one of the two pulses,
and after attenuating them to single photon level, reflects the
pulses back along the same path back to Bob. The pulses then travel
back through the AMZI, during which time Bob, modulates the other
pulse of the pair.  Such an arrangement automatically cancels any
slow variations in the difference of the two paths in the AMZI and
is now the basis of commercially available QKD systems
\cite{stucki02}.

Although the round-trip fibre architecture can reduce the problem of
phase drift, it can degrade the system performance. Contamination of
the quantum signal by photons from the strong pulse that are
back-scattered by the fibre contributes to the quantum bit error
rate (QBER). To minimise this, a round-trip system needs to operate
by sending bursts of pulses spaced by long dead intervals
\cite{stucki02}. This will reduce the duty cycle and bit rate of the
system. The round trip layout also cannot be used with a single
photon source. This is due not only to the effective doubling of the
transmission loss, but also because of the Trojan horse attack
\cite{gisin02}. Since Alice is simply a modulation station in such a
scheme, an eavesdropper could insert Trojan photons to be modulated
and gain information.

In this paper, we show that phase and polarization drifts in a
one-way QKD system may be controlled using active compensation.  In
this scheme, each signal pulse is multiplexed with brighter
reference pulses at the same wavelength.  Alice and Bob modulate
only the signal pulses, but leave the reference pulses un-modulated.
Therefore, the interference of the reference pulse is used to detect
any phase drift and provide a feedback signal to rebalance the
double AMZI. The system is self-initiating and can operate
continuously over long periods without user intervention.


\begin{figure*}[htbp]
\begin{center}
\includegraphics[width=15cm]{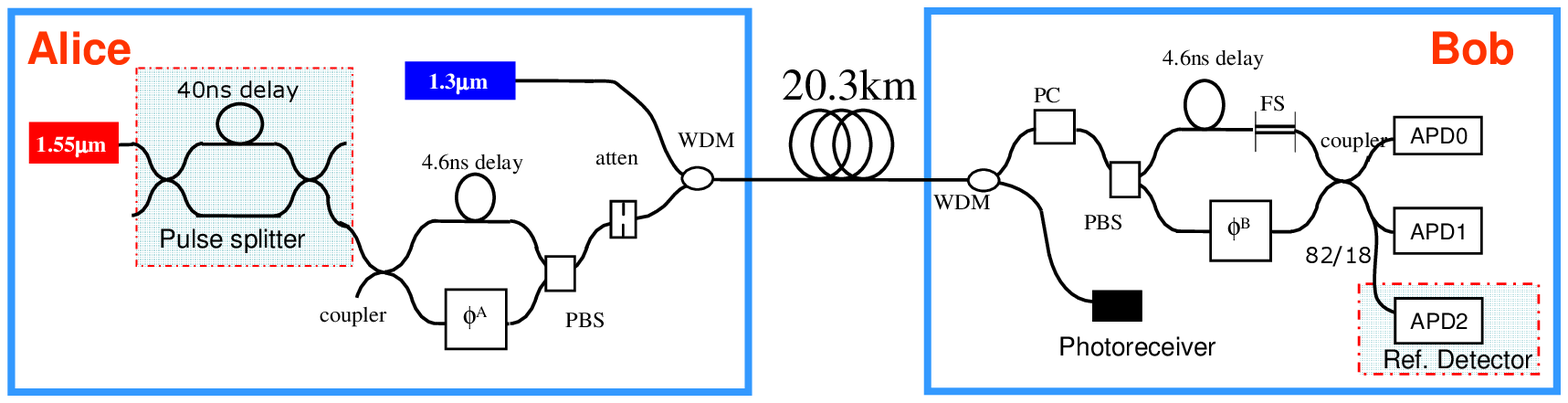}
\caption{ Schematic showing the fibre optic quantum key distribution
system with active phase \& polarisation compensation. WDM:
wavelength division multiplexer; FS: fibre stretcher; PBS:
polarising beam combiner/splitter; PC: polarisation controller;
$\phi^A$, $\phi^B$: phase modulator.}
\end{center}
\end{figure*}

Figure 1 illustrates a schematic of the optical layout in our
system. In comparison with the setup we reported previously to
achieve 122km QKD \cite{gobby04a}, we have added a pulse splitter
into Alice's arrangement and a reference detector at Bob's setup
(shown within the shaded inserts). The pulse splitter divides the
input 1.55~$\mu$m DFB laser pulse (of 400~ps duration) into two
pulses separated by a 40~ns delay. The reference (late) pulse is 24
times stronger than the signal (early) pulse. The output of the
splitter is fed into the encoding AMZI. The signal (early) pulse is
then attenuated to an average intensity of 0.2 photons per clock
cycle, before multiplexing with pulses from the 1.3~$\mu$m clock
laser, which serve as a timing reference. Bob's set-up contains
three InGaAs avalanche photodiodes (APD's); APD0 and APD1 for the
signal photons are cooled to -100$^o$C by a compact Stirling cooler
to achieve a dark count probability of $10^{-6}$, while APD2 for the
strong reference pulse, cooled by a thermal-electric cooler to
­40ºC, has a dark count probability of $2\times 10^{-5}$. The
overall detection efficiency of APD0 and APD1 is around 11\%. Their
detection rate is balanced because of the higher intrinsic
efficiency of APD1, despite of the 18\% splitting loss to APD2.
Active stabilization of the phase drift is achieved by using the
count rate in the reference detector APD2 to control the bias
applied to the fibre stretcher in Bob's AMZI.

The sending and receiving units were housed in compact 3U-height
19-inch rack mounts containing both the optics and associated
electronics. We found that the scheme could compensate for changes
in the ambient temperature inside the housing due to the drive
electronics.  The BB84 protocol \cite{bennett84} for QKD was used,
with the required classical information exchanged over the Ethernet.
The measurements results are sifted, and the QBER monitored, in
real-time. Bob also monitors the photon detection rate, the phase
drift/compensation rate, and the coincidence rate of APD0 and APD1.
The key distribution stations were connected using installed telecom
fibre serving our laboratory on the Cambridge Science Park.  Its
total length was measured to be 20.3 km, and its transmission loss
was 5.3dB for 1.55 m.  This is rather more than the specified
attenuation of the fibre (0.2 dB/km) due to connector and splice
losses.


\begin{figure}[htbp]
\begin{center}
\includegraphics[width=8cm]{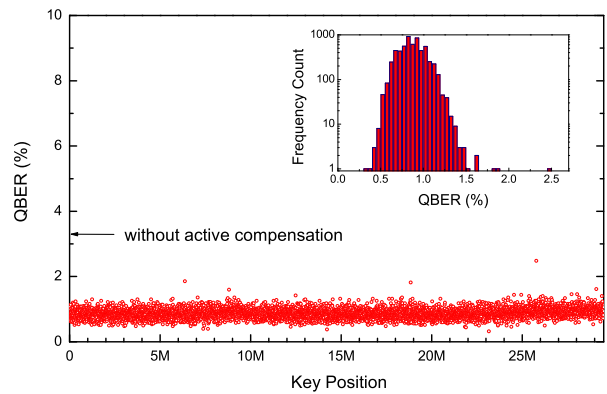}
\caption{ The measured quantum bit error rate (QBER) averaged over 5
kbits as a function of bin position. The inset shows the
distribution of different values for the QBER. The arrow indicates
the QBER obtained in the previous QKD system without active
compensation \cite{gobby04a,gobby04b}.}
\end{center}
\end{figure}

Figure 2 illustrates data recorded over an uninterrupted 19-hour
experimental run, during which 29.7~Mbit of sifted key material was
formed at an average rate of 0.43 kbits/s. This data was recorded
with a relatively low clock rate of 250kHz, in order to minimize the
APD afterpulse counts \cite{ribordy98} and thereby characterize
contributions of the active compensation scheme to the QBER. Figure
2 plots the QBER sampled over 5-kbits block-size as a function of
key position. Notice that most of the measured QBER lies between
0.5\% and 1.2\%, while the minimum value is just 0.32\%. The QBER
averaged over the whole QKD session is 0.87\%, which is among the
lowest reported values so far. The few points with slightly higher
QBER are related to times when the interferometer was highly
unstable due to the fibre-stretcher resets discussed below. However,
even during these periods, the averaged QBER remained less than
2.5\%, demonstrating that the system is able to recover quickly. The
probability of the QBER exceeding 1.5\% is low, as shown by the
distribution of measured values in the inset of Fig. 2.

\begin{figure}
\begin{center}
\includegraphics[width=8cm]{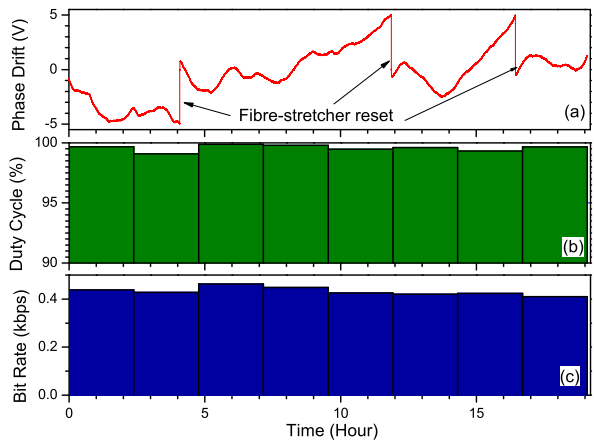}
\caption{(a) the recorded fibre-stretcher voltage, (b) the measured
duty cycle and (c) the measured bit rate as a function of time.}
\end{center}
\end{figure}

Figure 3(a) shows the voltage applied to the fibre stretcher as a
function of time. The drive voltage varies in the range from -5V to
5V.  With a phase compensation co-efficient of 294$^o$/V, the fibre
stretcher is able to give around a maximum of ~8-wavelength
compensation. When the phase drifts beyond this range, the stretcher
is programmed to reset to 0V. From the data collected, it is found
that the average phase compensation/drift rate is ~0.18$^o$/s. The
duty cycle of the system, \textit{ie}., the fraction of the time is
it actually distributing key material, is an important parameter for
assessing the performance of different QKD schemes.  The system may
sometimes encounter a short period of sharp phase drift it fails to
compensate, for example, during reset of the fibre stretcher voltage
or a sudden shock. As a result, it stops key sifting, and all
photons received during this period are wasted. The duty cycle   is
defined here as

\begin{equation}
\eta=\frac{2\ast sifted-key-size}{number-of-photons-received}
\end{equation}

Such a definition is equivalent to the ratio of the time that is
actually used for key formation to the total time. Figure 3(b) shows
the duty cycle as a function of time. It is always greater than
99\%, demonstrating that active phase compensation is highly
efficient. The average duty cycle over the 19-hour period is 99.6\%.
The system also corrects for polarisation drift in the fibre. This
is compensated by walking the driving voltages upon the three axes
of the polarization controller at a fixed step-size. The sum of the
count rates in APD0 and APD1 is used as a feedback signal to the
controller.  The bit rate remains relatively stable across the QKD
session, as shown in Fig. 3(c), indicating that this simply strategy
is effective.

There are three significant contributions to the QBER: imperfections
of the optical layout, detector erroneous counts, and phase
mis-compensation. Of these, imperfection of the double AMZI, due to,
for example, misalignment of the polarization maintaining components
or non-ideal splitting ratios for the couplers, makes the largest
contribution.  This is illustrated by the fact that the classical
interference fringe visibility was recorded to be 99.12\%,
suggesting the QBER due to the optical imperfection is 0.44\%. This
error source can be reduced to be less than 0.1\% in an optimized
double AMZI [5].

\begin{figure}
\begin{center}
\includegraphics[width=8.8cm]{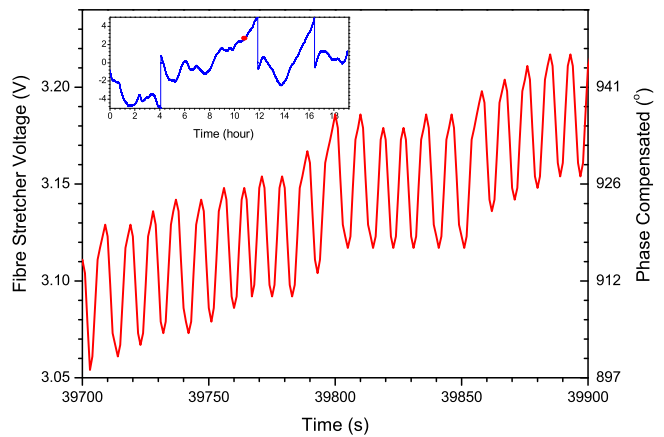}
\caption{A short section of the recorded fibre stretcher voltage and
the corresponding phase compensated as a function of time. }
\end{center}
\end{figure}

The QBER due to the residual mismatch in the phase delay between the
two AMZIs is an important parameter to evaluate the success of the
active compensation scheme. Figure 4 shows the same data as in Fig.
3(a), but over a much shorter timescale.  Notice that the voltage
applied to the fibre stretcher oscillates around an average value
with an amplitude of 70~mV, which corresponds to maximum error in
the phase compensation of  10.3$^o$. The QBER due to
mis-compensation can be approximated by

\begin{equation}
QBER_{phase}=\int_{-\frac{\Delta \varphi}{2}}^{+\frac{\Delta
\varphi}{2}}\frac {1-\cos \varphi}{2}d \varphi
\end{equation}

\noindent where $\Delta \varphi$ is the phase variation range due to
mis-compensation. Using the recorded data, phase mis-compensation is
found to contribute only 0.27\% to the QBER. The remaining
contribution to the QBER, of around 0.15\%, is attributed to the
erroneous counts in the single photon detectors, including dark
counts, stray photons and after-pulsing.

We have also tested the system at a clock frequency of 1MHz, and
obtained a sifted bit rate of 1.7 kbit/s. However, the average QBER
was found to increase to 1.8\%, due to an increase in afterpulse
noise in the APDs.  We stress that this increase in the QBER is due
to the APDs and is not a limitation of the active compensation
scheme.  Given a sufficiently improved APD \cite{yoshizawa04}, clock
rates in excess of 10~MHz may be expected. Finally, we point out
that the active compensation scheme is compatible with a single
photon source by only a slight modification to the set-up. Single
photons can be combined with reference pulses from an attenuated
laser using a highly asymmetric coupler. Alternatively, they can be
multiplexed by shining the reference laser pulse through the lower
input port of Alice's encoding AMZI.

In summary, we present an automated and compact QKD system with
active compensation for both polarization and phase drifts. The
compensation scheme gives a QBER of 0.87\% averaged over 19 hours
key distribution session with an averaged duty cycle of 99.6\%.  The
results demonstrate that actively compensated QKD systems are
suitable for practical applications.

\end{document}